\documentclass[preprint,
superscriptaddress,
 amsmath,amssymb,
 aps,
 prl,
]{revtex4-2}
\usepackage{textcomp,mathcomp}
\usepackage{graphicx}% Include figure files
\usepackage[utf8x]{inputenc}
\usepackage{dcolumn}% Align table columns on decimal point
\usepackage{bm}% bold math
\usepackage{xcolor}
\usepackage{caption}
\usepackage[labelformat=simple]{subcaption}

\usepackage{pgfplots}
\pgfplotsset{width=\columnwidth,compat=1.8}
\usepackage{tikz}
\usepackage{textgreek}
\usepackage{soul}

\begin{document}

%\preprint{APS/123-QED}

\title{High Yield of Alpha Particles Generated in Proton-Boron Nuclear Fusion Reactions Induced in Solid Boron Hydride B${}_{18}$H${}_{22}$}% Force line breaks with \\

\author{M. Kr\r us}
    \email{krus@ipp.cas.cz}
 \affiliation{Institute of Plasma Physics of the Czech  Academy of Sciences, Za Slovankou 1782/3, 182 00 Prague, Czechia}%Lines break automatically or can be forced with \\
\author{M. Ehn}%
\affiliation{Institute of Inorganic Chemistry of the Czech  Academy of Sciences, 250 68 Husinec-\v Re\v z, Czechia}% 
\author{M. Kozlov\' a}
\affiliation{Institute of Plasma Physics of the Czech  Academy of Sciences, Za Slovankou 1782/3, 182 00 Prague, Czechia}%
\author{J. Bould}%
\affiliation{Institute of Inorganic Chemistry of the Czech  Academy of Sciences, 250 68 Husinec-\v Re\v z, Czechia}%
\author{P. Pokorn\' y}
\affiliation{Institute of Plasma Physics of the Czech  Academy of Sciences, Za Slovankou 1782/3, 182 00 Prague, Czechia}%
\affiliation{Czech Technical University in Prague, Faculty of Nuclear Sciences and Physical Engineering, B\v rehov\' a 7, 115 19 Prague, Czechia}
\author{P. Gajdo\v s}
\affiliation{Institute of Plasma Physics of the Czech  Academy of Sciences, Za Slovankou 1782/3, 182 00 Prague, Czechia}%
\affiliation{Czech Technical University in Prague, Faculty of Nuclear Sciences and Physical Engineering, B\v rehov\' a 7, 115 19 Prague, Czechia}
\author{R. Dud\v z\' ak}
\affiliation{Institute of Plasma Physics of the Czech  Academy of Sciences, Za Slovankou 1782/3, 182 00 Prague, Czechia}%
\affiliation{Institute of Physics of the Czech  Academy of Sciences, Na Slovance 1999/2, 182 00 Prague, Czechia}%
\author{O. Renner}
\affiliation{Institute of Plasma Physics of the Czech  Academy of Sciences, Za Slovankou 1782/3, 182 00 Prague, Czechia}%
\author{M. Guldan}
\affiliation{Czech Technical University in Prague, Faculty of Nuclear Sciences and Physical Engineering, B\v rehov\' a 7, 115 19 Prague, Czechia}
\author{M. My\v ska}
\affiliation{Czech Technical University in Prague, Faculty of Nuclear Sciences and Physical Engineering, B\v rehov\' a 7, 115 19 Prague, Czechia}
\author{L. \v Skoda}
\affiliation{Czech Technical University in Prague, Faculty of Nuclear Sciences and Physical Engineering, B\v rehov\' a 7, 115 19 Prague, Czechia}
\author{M. G. S. Londesborough}
 \email{michaell@iic.cas.cz}%
\affiliation{Institute of Inorganic Chemistry of the Czech  Academy of Sciences, 250 68 Husinec-\v Re\v z, Czechia}%

%\affiliation{Institute of Physics of the CAS, Na Slovance 1999/2, 182 00 Prague, Czechia}%

\date{\today}% It is always \today, today,

\begin{abstract}
The use of solid boron hydride molecules as a fuel for proton-boron fusion was proposed by M. Krus and M. Londesborough at "Interaction of Inorganic Clusters, Cages, and Containers with Light" workshop in November 2021. Here we demonstrate experimentally, for first time, that the solid boron hydride, octadecaborane - \textit{anti}-B${}_{18}$H${}_{22}$, produces a relatively high yield of alpha particles of about $10^9$ per steradian using a sub-nanosecond, low-contrast laser  pulse (PALS) with a typical intensity of $10^{16}$ Wcm${}^{-2}$. In contrast to previously published proton-boron studies, the boron hydrides, due to their composition containing only atoms of boron and hydrogen, represent a natural choice for future proton-boron nuclear fusion schemes. 

\end{abstract}

\maketitle

%\tableofcontents

\section{\label{sec:level1}Introduction}

The nuclear reaction of proton and boron-11 nuclei (pB) releasing three alpha particles was discovered by Mark Oliphant \cite{doi:10.1098/rspa.1933.0117} in the Cavendish laboratory in 1933.
\begin{equation}
    p + {}^{11}B \rightarrow 3 \alpha
    \label{equat1}
\end{equation}
Since that time, this reaction has been extensively studied to determine its cross section (or fusion rate) both experimentally and theoretically \cite{QUEBERT1969646, Dmitriev2009, osti_4014032, DAVIDSON1979253,Becker1987, PhysRev.139.B818, STAVE201126, Spraker2012, Sikora2016, LIU2002107}, as well as to assess alpha particle angular distributions \cite{SYMONS196393, PhysRev.91.606, Kamke1967}. Such a reaction was later proposed as a source of fusion energy. However, calculations have suggested that a conventional confinement fusion scheme is difficult due to the large bremsstrahlung radiation losses and alpha particle energy deposition in the plasma \cite{Moreau_1977}. As an alternative, several non-thermal schemes have been proposed for fusion ignition and burning, such as a heat detonation wave \cite{MARTINEZVAL1996142} and plasma block collision \cite{Hora81, hora_2015}.   Computational analysis of these schemes predicted fusion rates that indicate the possibility for successful pB fusion \cite{W.M.Nevins_2000, Krainov_2005}. This recognition has sparked an increasing interest in the pB reaction as a viable alternative for a clean neutron-less fusion energy source and as a means to enable the pain-free and non-invasive hadron therapy of cancer through the generation of alpha particles that augment the radiation effect \cite{Yoon2014, Cirrone2018, Giuffrida2016, Blaha2021}. Another promising potential application of such an alpha particle source is the production of radioisotopes for imaging therapy or radiotracing \cite{Qaim2016, SZKLINIARZ2016182}. Moreover, the high energy gamma photons, which can also be emitted (with negligible cross section \cite{Hegelich}) during pB reactions, might be used for online proton beam monitoring during cancer treatment \cite{Petringa_2017_1}.
%, Petringa_2017_2}. 

Recently, rapid advances in laser technology have opened new approaches to laser-ignition of pB fusion.  A systematic study of these alternatives began in 2005 by application of a high-power laser to initiate pB fusion reactions in boron-enriched polyethylene \cite{PhysRevE.72.026406}. Within the scope of this study, experiments detected around 1000 fusion alpha particles per laser pulse. Eventually, this result was corrected to the yield of $10^5$ alpha particles per steradian per shot. Subsequently, experiments performed at LULI by combining two laser pulses - a nanosecond pulse for the ionization of a solid pure boron target, and a picosecond pulse for proton acceleration via the TNSA (target normal sheath acceleration) mechanism \cite{10.1063/1.1333697} resulted in pB fusion with an alpha particle yield of $10^7$\,\textalpha/sr/shot \cite{Labaune2013}.  

These pioneering experiments engendered further studies.  For example, pB fusion was successfully achieved with a yield of $10^{10}$ \textalpha/sr/shot in 2020 at PALS laser \cite{PhysRevE.101.013204}, using boron nitride targets containing hydrogen. Other experimental schemes can be divided into two main groups: (i) the ``pitcher-catcher'' scheme, using an accelerated proton beam  focused onto a pure boron target \cite{Labaune2013, bonvalet2021, Tayyab_2019, baccou2015} or boron nitride target \cite{margaronefrontiers}, and (ii) the ``in-target'' scheme, using the direct irradiation of a boron-doped polyethylene target \cite{PhysRevE.72.026406}, a boron nitride target \cite{PhysRevE.101.013204}, or a boron-doped hydrogen-enriched silicon target \cite{Margarone_2014, Picciotto_2014}.

In addition to these laser triggered pB fusion schemes, alternative methods are currently under investigation. Examples are colliding beam fusion \cite{Rostoker_1997} using a Field-Reversed Configuration \cite{Tuszewski_1988} and, more recently, pB fusion was demonstrated at the Large Helical Device stellarator \cite{Magee_2023}, being the first realisation of pB fusion in a magnetically confined plasma.

The laser pulse driven generation of alpha particles by pB fusion (in-target geometry) was recently demonstrated at a kHz repetition rate using a 10\,GW, 20\,mJ laser system focused on boron nitride targets with a thin polymer film layer as a source of protons. The yield of alpha particles reached $10^{6}$ alpha particles per second \cite{Istokskaia2023}.

Despite these early achievements, the search for an optimal pB fuel remains in its infancy and yet it is a key aspect for the ultimate success of applicable pB fusion.  Hitherto, all tested targets comprised compounds of boron with heteroelements such as nitrogen, silicon, or carbon-rich polymer films, offering only sparse boron content and the multi-element contamination interferes with the pB fusion process by diverting energy from the system.  It is, therefore, of interest to investigate the potential of a group of compounds containing only atoms of hydrogen and boron - the boron hydrides (commonly referred to as the boranes) as fuels for the pB fusion process.   The boranes are a broad family of spherically-aromatic molecules with polyhedral cluster geometries of atoms of boron surrounded by a sheath of hydrogen atoms. They do not occur naturally, but are readily synthesized in specialised laboratories from abundantly available natural materials, where multi-gram and kilogram scale production is feasible. 
 The spatial separation between boron and hydrogen atoms is a chemical bond extending over approximately 0.1\,nm.  The lighter boranes, such as diborane (B${}_{2}$H${}_{6}$) and pentaborane (B${}_{5}$H${}_{9}$), are gases.  However, increasing molecular weight leads to higher boranes, such as decaborane (B${}_{10}$H${}_{14}$) and octadecaborane (B${}_{18}$H${}_{22}$), which are air-stable crystalline solids.  These borane materials thus offer dense concentrations of hydrogen and boron atoms in the correct ratios for fusion whilst being devoid of any other element that may hinder the pB reaction.   

Within this work, we provide the first demonstration of the use of solid boranes as a pB fuel (within in-target geometry) and show the high yield generation of alpha particles by sub-nanosecond kJ-class laser system irradiation of solid B${}_{18}$H${}_{22}$ targets.      

\section{\label{sec:level2}Experimental setup}
The experimental campaign was performed in Prague at the PALS laser facility utilizing an iodine photodissotiation laser carrying 550\,J in 350\,ps (FWHM) pulses at a fundamental wavelength of 1315\,nm \cite{JungwirthPALS}.{} The laser beam was tightly focused onto the boron hydride target by a f/2 lens at zero degrees with respect to the target normal axis. The focal spot, with a diameter of about 80\,\textmu m, implies that the laser intensity at the target reached about $10^{16}$\,W/cm${}^2$.

Compressed boron hydride targets (octadecaborane - \textit{anti}-B${}_{18}$H${}_{22}$ - see Figure 1) with a density of 0.9\,g/cm${}^3$ were irradiated by the laser to trigger the proton-boron fusion reaction. The octadecaborane target, containing a natural mixture of boron isotopes - 20\,\% of ${}^{10}$B and 80\,\% of ${}^{11}$B, was synthesized at the Institute of Inorganic Chemistry using methods based on previously published procedures \cite{YLi}.  Pure octadecaborane targets each contained 150 mg of material compressed by 7.5t of pressure into discs of 13\,mm diameter and 1.2\,mm thickness.

\begin{figure}
    \centering
    \includegraphics[width=1\linewidth]{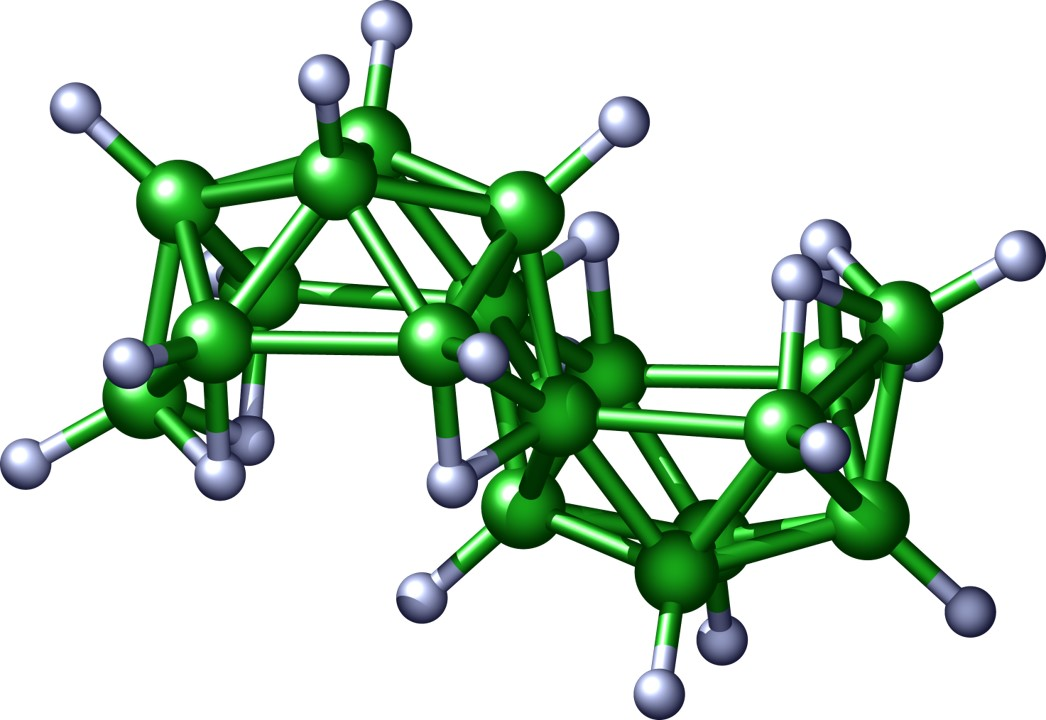}
    \caption{Molecular structure of \textit{anti}-B${}_{18}$H${}_{22}$. Green atoms are boron, white atoms are hydrogen.}
    \label{fig:enter-label}
\end{figure}

B${}_{18}$H${}_{22}$ was chosen amongst other molecular boranes for these initial experiments because of its chemical stability, relative non-volatility, appropriate B:H ratio (based on the stoichiometry of Equation \ref{equat1}), and its propitious behaviour in electromagnetic fields \cite{Mich1, Mich2, Mich3}.  

\begin{figure}
    \centering
    \includegraphics[width=1\linewidth]{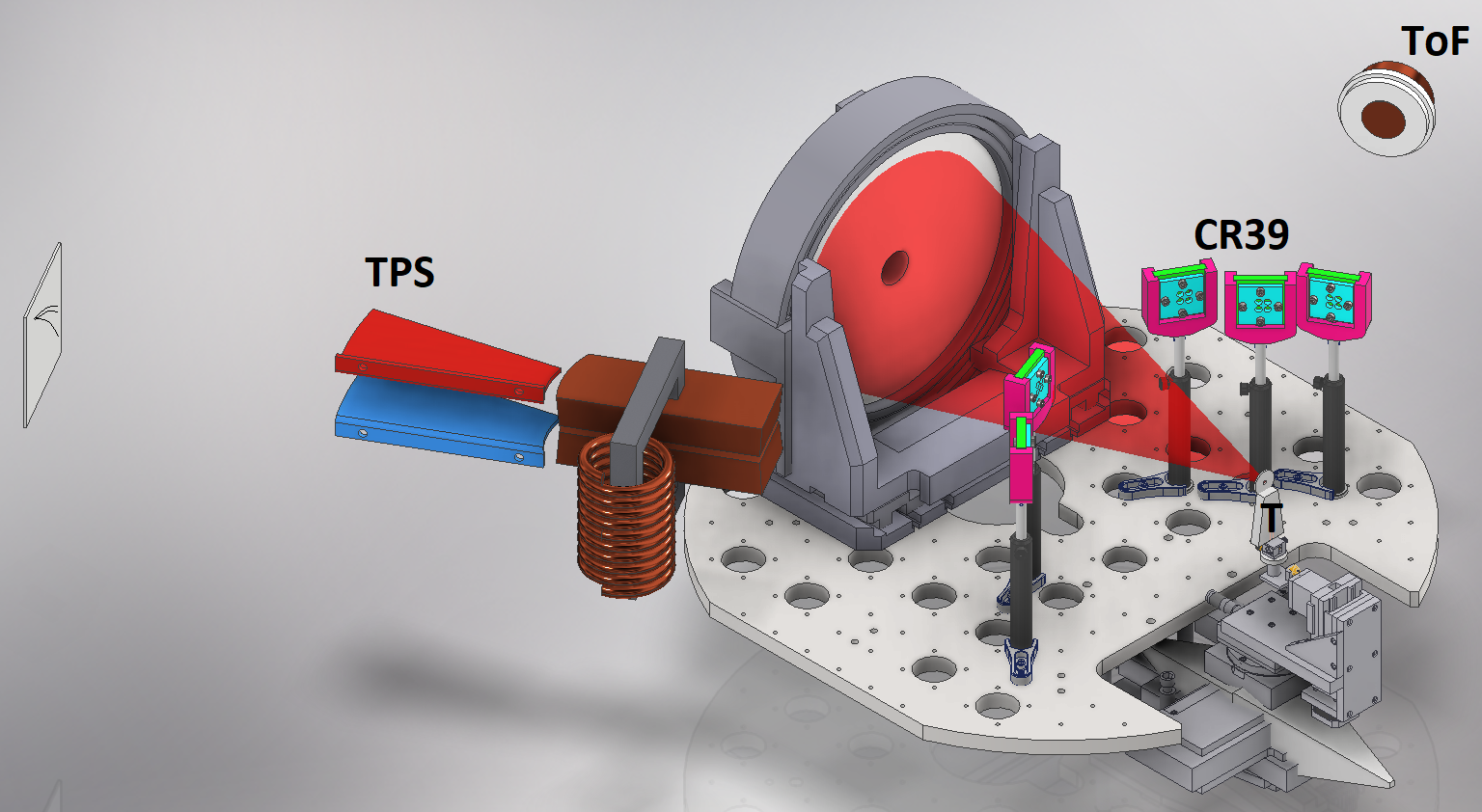}
    \caption{Schematic drawing of the experimental setup. Laser beam (red cone) is focused (f/2 lens) on octadecaborane target (T). The main diagnostics for alpha particles are an array of five CR39 nuclear track detectors, a Thomson parabola  spectrometer (TPS) comprising parallel magnetic (electromagnet) and electric fields, and Time-of-Flight spectrometer (ToF).}
    \label{fig:exp_setup}
\end{figure}

Within this experiment, all the detectors occupied the backward half space of the vacuum chamber (with respect to the direction of the laser beam propagation). The experimental setup is shown in Figure \ref{fig:exp_setup}.

The main diagnostics used in this experiment to monitor alpha particle production was a set of CR39 (PADC - polyallyl diglycol carbonate) nuclear track detectors. These CR39 detectors enable the shot-to-shot measurement of the absolute number of the generated alpha particles. Five detectors were used in the chamber and positioned to optimise the resolution of alpha-particle detection. Three of the detectors were placed at larger angles to the incident beam - 60\,deg, 70\,deg and 80\,deg - where the direct emission of high energy plasma particles is relatively low whereas the alpha particle emission is isotropic. Two further CR39 detectors were positioned near the laser axis (at angles of 40\,deg and 50\,deg) to monitor the plasma particle flux. Each detector is divided into four sectors of the same size; three of these sectors were covered by a thin aluminum foil (a thickness of 5\,\textmu m, 10\,\textmu m and 20\,\textmu m), and one sector without foil. The aluminum filters stop low energy plasma ions and thus reduce particle flux in order to clearly distinguish adjacent particle tracks (i.e. minimizing the track trace overlap).

A second diagnostic tool, a Thomson parabola spectrometer (TPS), with a microchannel plate sensor, was installed at an angle of 30 degrees with respect to the target normal thus enabling the monitoring of the energy of particles emitted from the target.  This distinguishes between the different ion species present in the generated plasma. Furthermore, TPS is capable of monitoring shot-to-shot ion energy distribution and proton and boron ion cut-off energy.

Lastly, a time-of-flight (ToF) spectrometer, equipped by an ion collector detector with a collecting area diameter of 20\,mm, was positioned at an angle of 60\,degrees with respect to the target normal axis.  This was used for shot-to-shot measurements of the energy distribution of protons and alpha particles (lower energy particles and heavier ions were filtered out by a 2\textmu m thin aluminum foil located in front of the ion collector).

\section{\label{sec:level3}Results}
\begin{figure}[htp]
\centering
\begin{tabular}{@{}c@{}}
\subfloat{\includegraphics[width=\linewidth]{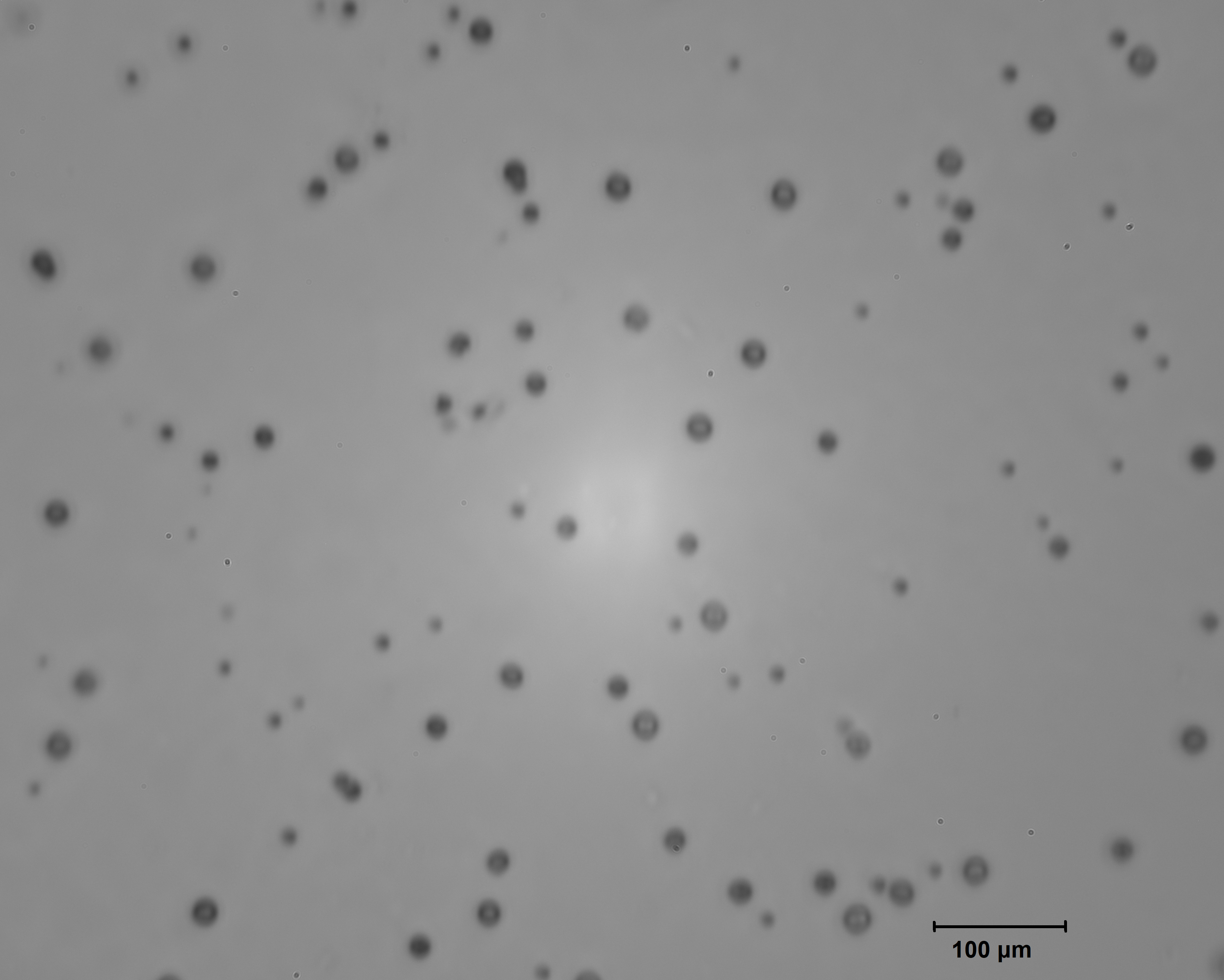}}\\ (a)
\end{tabular}\qquad % some space
\begin{tabular}{@{}c@{}}
\subfloat{\includegraphics[width=\linewidth]{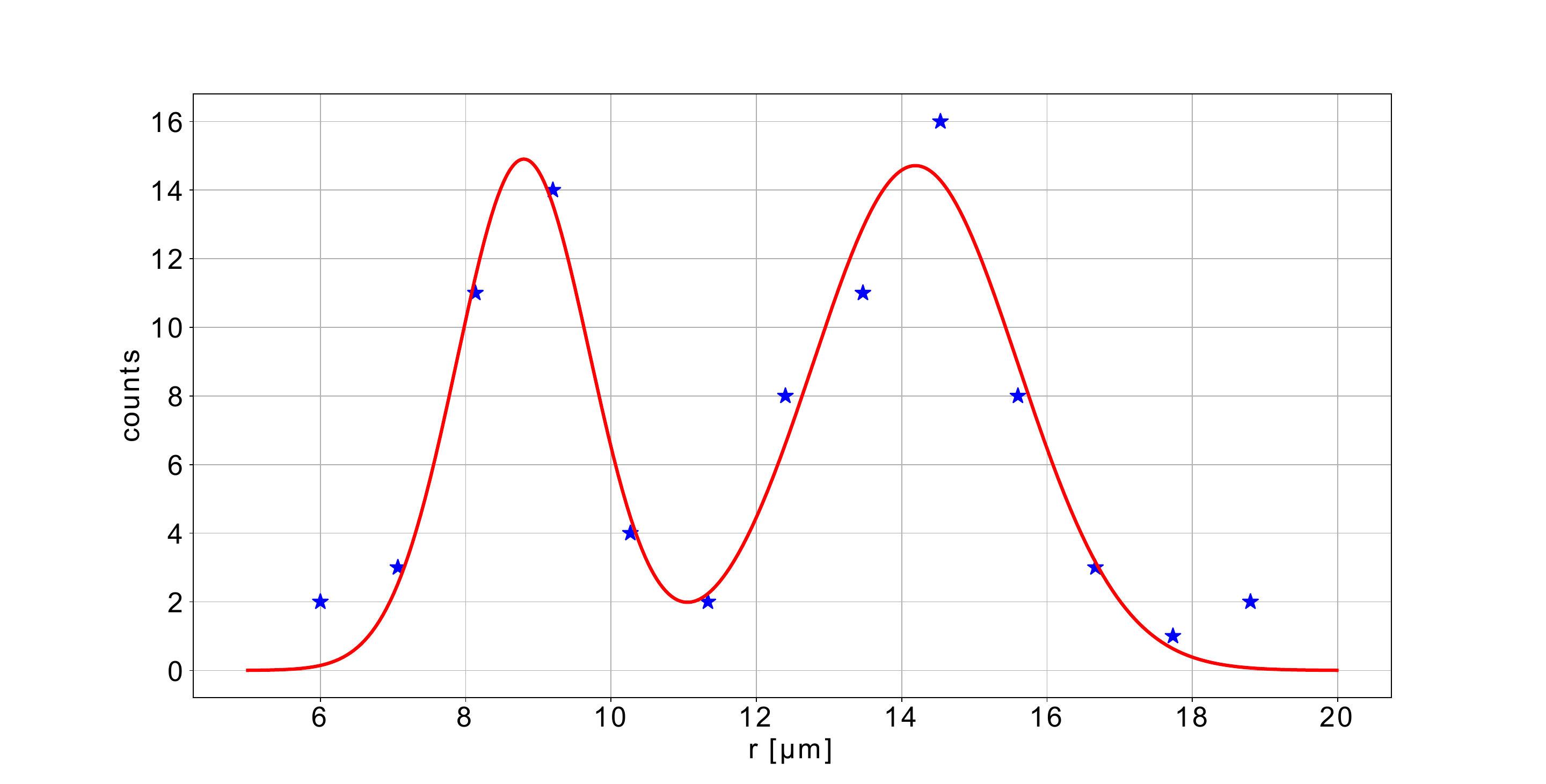}}\\ (b)
\end{tabular}\qquad % some space
\caption{a) Typical microscope image of visualized ion track in etched CR39 nuclear track detector. Smaller spots correspond to proton and larger spots to alpha particle tracks. For better understanding, this image was taken in the marginal part of the sensitive area in which the total number of tracks is smaller. b) Histogram (blue stars, red curve represents double Gaussian fit) of analyzed etched track radii (obtained from above image) clearly distinguishes the proton tracks (with 9\,$\mu$m in radius) and alpha particle tracks (with 14\,\textmu m in radius).}
 \label{fig:CR39image}
\end{figure}

The primary diagnostic tool capable of identifying alpha particles with the highest relative certainty are the CR39 detectors, which also assess the total number of  alpha particles generated. In our experiments, the CR39 detectors were removed after each shot/exposure and chemically developed by immersion for 45 minutes in a 6M NaOH aqueous solution maintained at a temperature of 70\,\textcelsius. This process reveals a 2-dimensional cross-section of particle tracks, seen as spots, that were then examined by an optical microscope (with a magnification of 100) coupled with a pixel (CMOS) sensor. A typical image is shown in Figure \ref{fig:CR39image}; the smaller spots are created by tracks that may be assigned to protons, the larger spots are attributed to alpha particles. The distribution of small (proton) and large (alpha) spots are depicted in Figure 3b.

In order to verify the assumption that the large spots in Figure \ref{fig:CR39image} are indeed created by alpha particles we performed two reference experiments. The first one used a  polyethylene foil irradiated by the laser under the same conditions as the octadecaborane target. In this case, only spots with smaller sizes corresponding to proton tracks were observed (see Figure S1 in Supporting Information). In the second reference experiment, the CR39 detectors were irradiated by alpha particles emitted from an ${}^{241}$Am radioactive source via $\alpha$-decay, having almost identical energies as expected fusion-produced alpha particles. Here, the track sizes observed  were very similar to those corresponding to the alpha particles in Figure \ref{fig:CR39image} (see Figure S2 in Supporting Information). Together, these observations suggest with a high degree of certainty the generation of alpha particles during the laser pulse-octadecaborane interaction. Extrapolation of the CR detector signal data over a full solid angle (i.e. a complete spherical collection) was used to approximate the total number of produced alpha particles and thus a yield of alpha particles of $1.7\times10^9$/sr/shot.

\begin{figure}
    \centering
    \includegraphics[width=1\linewidth]{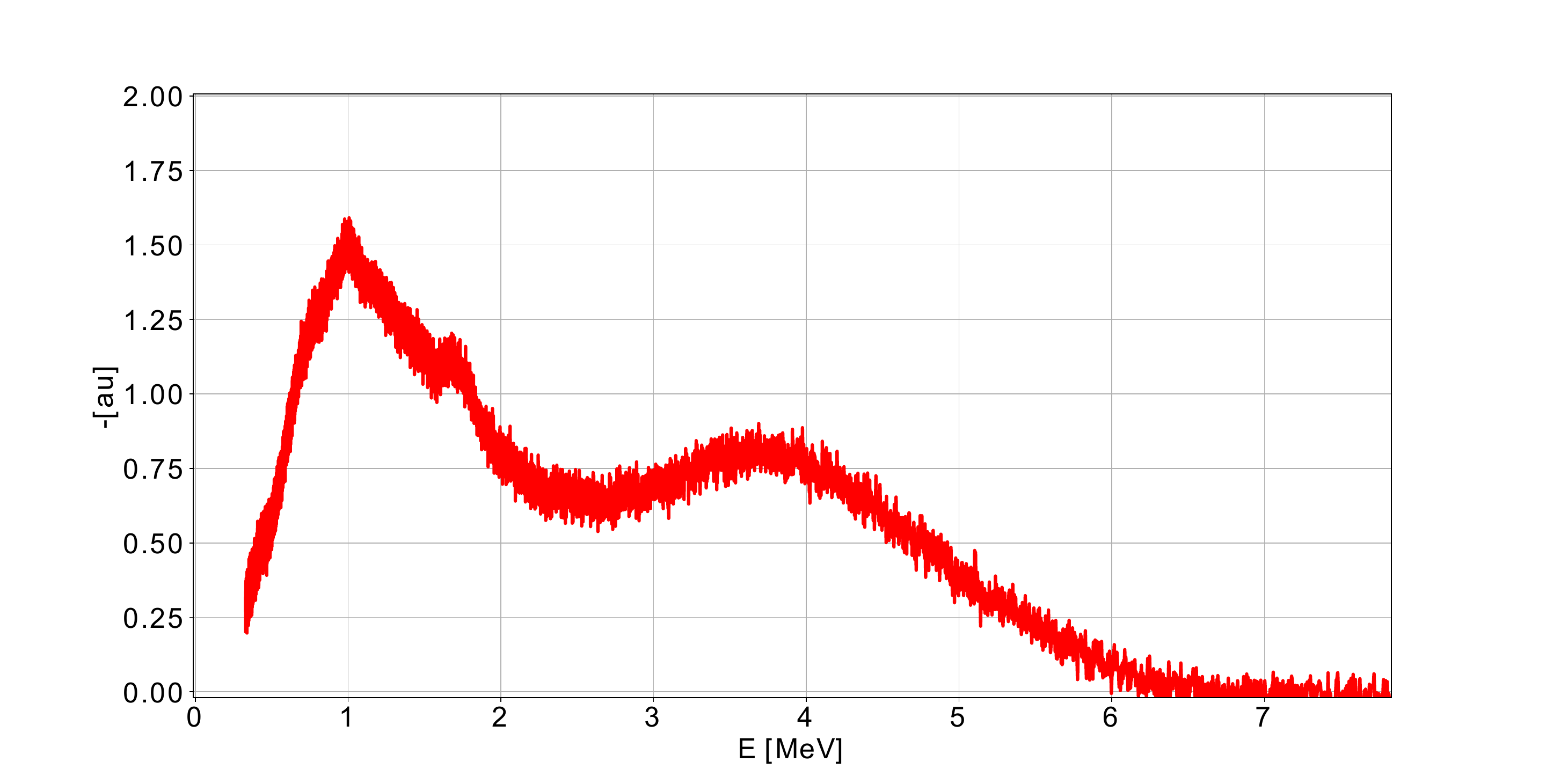}
    \caption{Time of flight spectrum recorded by oscilloscope connected to an ion collector. The spectrum shows a peak around 3.8\,MeV corresponding to the fusion-generated alpha particle energy.}
    \label{fig:tof_spectrum}
\end{figure}

Additional evidence for alpha particle generation in our experiment was provided by the ToF spectrum, which contains a peak at around 3.8\,MeV  (see Figure \ref{fig:tof_spectrum}) that corresponds well to the expected energy of 3.76\,MeV for alpha particles resulting from nuclear fusion. Although the low energy region ($<$ 2\,MeV) of the spectrum is complicated by electromagnetic interference induced in the ion collector circuit and cables, the peak corresponding to the alpha particle emission is clearly visible in the spectrum.

\begin{figure}[htp]
\centering
\begin{tabular}{@{}c@{}}
\subfloat{\includegraphics[width=\linewidth]{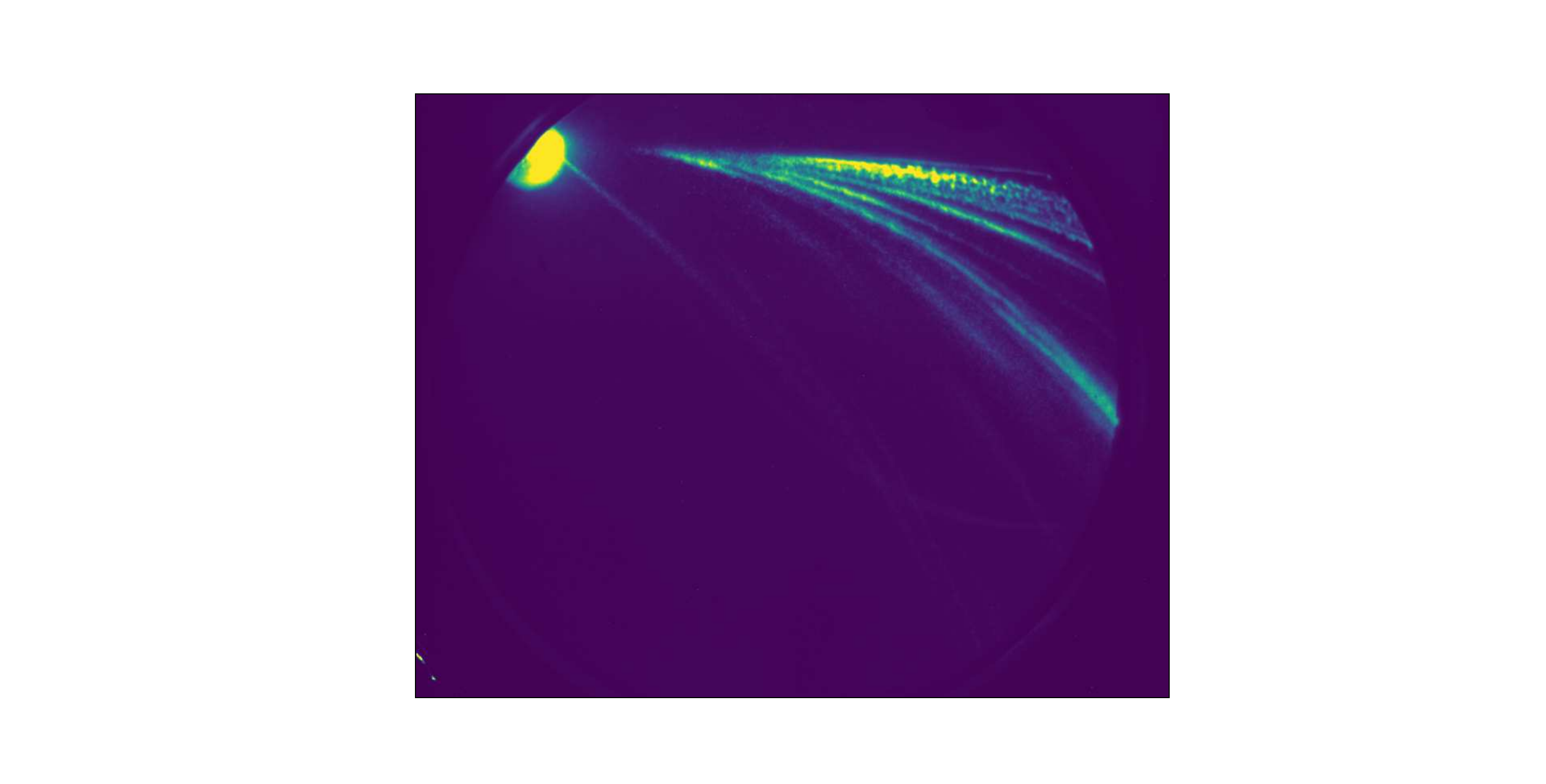}}\\ (a)
\end{tabular}\qquad % some space
\begin{tabular}{@{}c@{}}
\subfloat{\includegraphics[width=\linewidth]{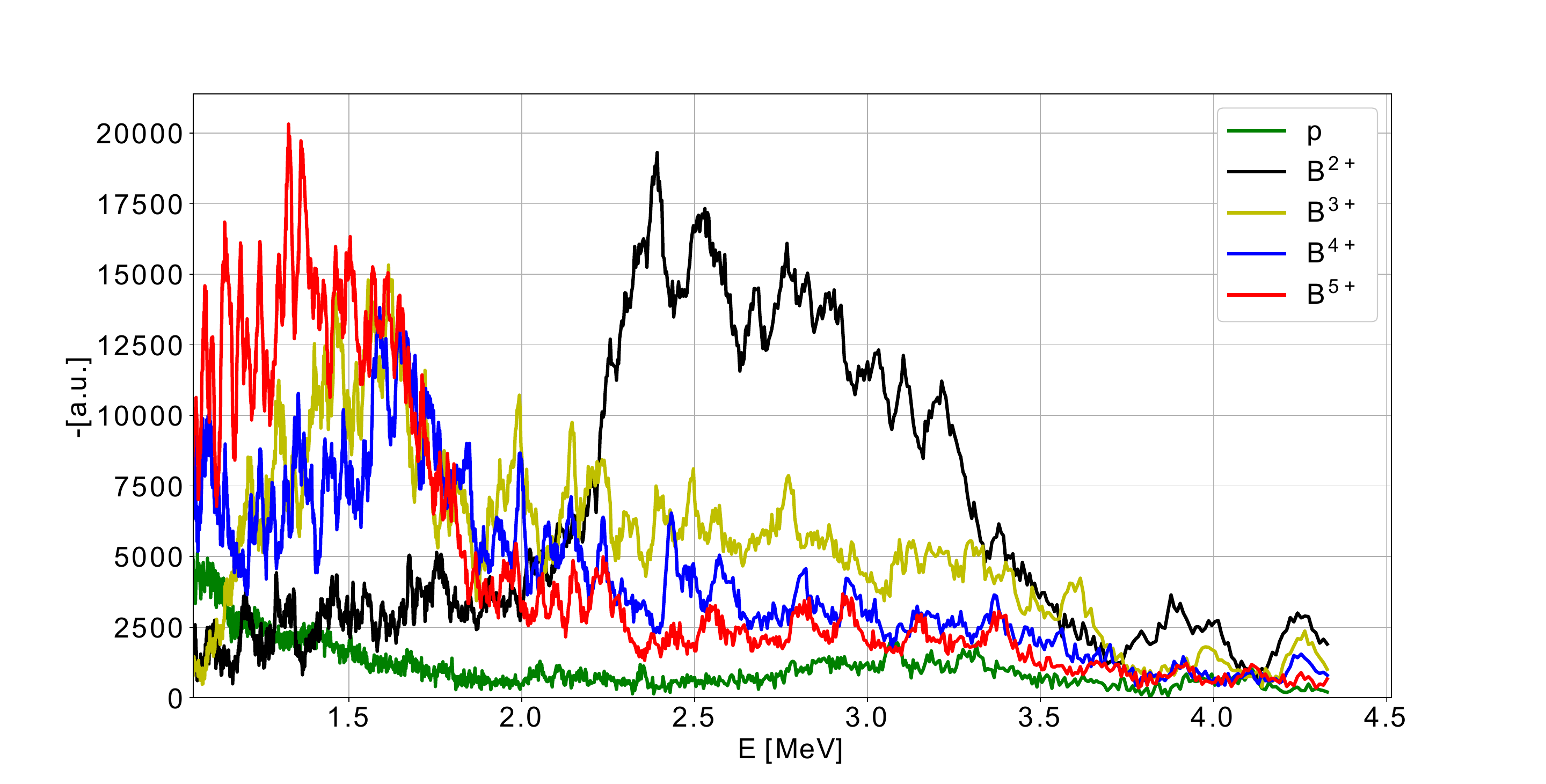}}\\ (b)
\end{tabular}\qquad % some space
\caption{a) Typical image of proton and boron ion parabolae. b) Proton and boron ion energy distributions obtained along the particular parabola. (Energy distribution of B${}^{+}$ is not shown as the amount of these ions was negligible in the observed energy range.)}
 \label{fig:TP_spectrum}
\end{figure}

TPS used to monitor the plasma ion energies produced in the experiment and thereby assess whether they meet the conditions for high cross section alpha particle production (i.e. $> 0.67$\,MeV) are depicted in Figure \ref{fig:TP_spectrum}. These data indicate that there are a large number of ions emitted from the created plasma with the sufficient energy to trigger pB fusion. Furthermore, the enhanced signal of particles with energies $>2$\,MeV for all ion states suggest the action of non-linear effects during the laser-plasma interaction.  These high-energy particles, however, also contribute to the pB fusion reaction.

\section{\label{sec:level4}Conclusion}
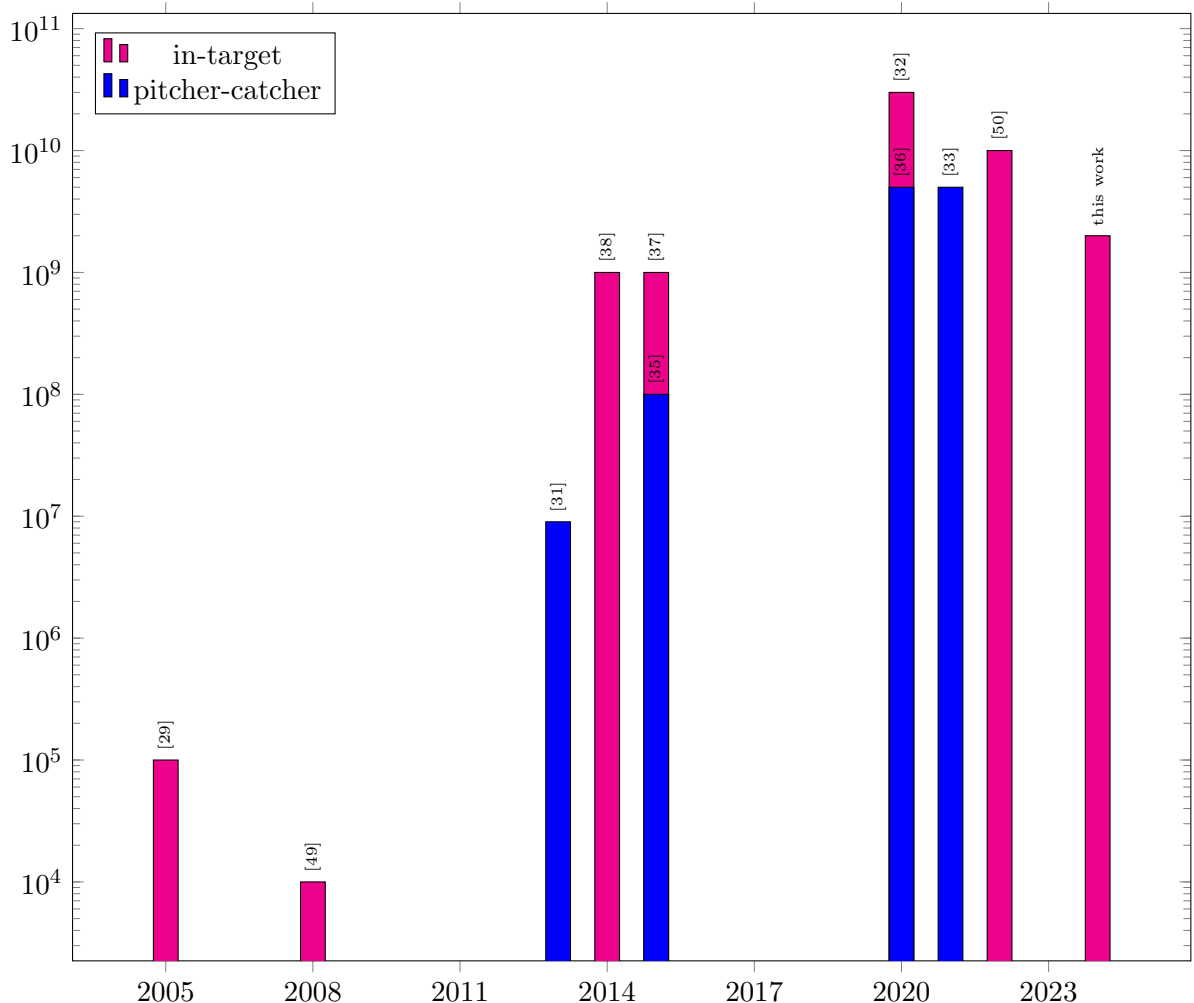
\begin{figure}
    \centering

\pgfplotstableread[row sep=\\,col sep=&]{
    interval & carT & carD & refer \\
     2005    & 100000  &   & \cite{PhysRevE.72.026406}\\
    2008     & 10000 &   & \cite{bonasera}\\
    2013    &  & 9000000 & \cite{Labaune2013}\\
    2014   & 1000000000 &  & \cite{Picciotto_2014}\\
    2015   & 1000000000  &  & \cite{Margarone_2014}\\
    2015 & & 100000000 & \cite{baccou2015}\\
    2020      & 30000000000  &  & \cite{PhysRevE.101.013204}\\
    2020 & & 5000000000 & \cite{margaronefrontiers}\\
    2021 &  & 5000000000 & \cite{bonvalet2021}\\
    2022 & 10000000000 & & \cite{margarone22}\\
    2024 & 2000000000 & & this work\\
    }\mydata

\begin{tikzpicture}

    \begin{axis}[
            ybar,
            ymode=log,
            bar width = 0.02\columnwidth,
            bar shift=0mm,
           symbolic x coords={2005, 2006, 2007, 2008, 2009, 2010, 2011, 2012, 2013, 2014, 2015, 2016, 2017, 2018, 2019, 2020, 2021, 2022, 2023, 2024},
            xtick={2005, 2008, 2011, 2014,2017, 2020, 2023},
            point meta=explicit symbolic,
            legend style={at={(0.02,0.98)},anchor=north west},
            nodes near coords,
            nodes near coords align={vertical},
            nodes near coords style={font=\tiny},  
            every node near coord/.append style={rotate=90, anchor=west},
        ]
        \addplot[fill=magenta, text=black] table[x=interval,y=carT,meta=refer]{\mydata};
        \addplot[fill=blue, text=black] table[x=interval,y=carD,meta=refer]{\mydata};
    \legend{in-target, pitcher-catcher}
    
    \end{axis}
\end{tikzpicture}
    \caption{Experimental results on laser-driven proton-boron fusion which were performed over last two decades. The alpha-particle yield achieved in the presented experiment is comparable to yield obtained both from in-target and pitcher-catcher fusion scheme.}
    \label{fig:barchart}
\end{figure}

This research marks a significant step forward in the exploration of boron hydrides as a highly promising fuel for laser-driven proton-boron fusion. Initial experiments have revealed octadecaborane (B${}_{18}$H${}_{22}$) to exhibit remarkable potential, demonstrating alpha particle production at yields ($1.7\times10{}^9$/sr/shot) comparable to the most exceptional results achieved thus far (see Figure \ref{fig:barchart}). This success instills confidence in the prospect of further investigations with these compounds, which boast extensive structural diversity. It is anticipated that continued research efforts will lead to substantial progress in realizing the efficient utilization of proton-boron fusion as a viable energy source.

\section{acknowledgements}
\begin{acknowledgments}

This work was performed jointly at the Institute of Inorganic Chemistry of the Czech Republic and at the PALS Research Infrastructure supported by a program of Ministry of Education, Youth and Sports of the Czech Republic (project No. LM2023068). This research is partially funded by Czech Science Agency, project No. GA23-07563S. This project has received funding from the European Union’s Horizon Europe research and innovation programme under grant agreement no 101096317. 
%UK participants in the Horizon Europe Project V4F are supported by UKRI grant number 10062154 (MODUS).
The authors thank the technical personnel of PALS Research Facility for their support during the experimental campaign.
\end{acknowledgments}

\bibliography{krus_preprint}% Produces the bibliography via BibTeX.

\end{document}